\documentclass[10pt]{article}
\usepackage{times}
\usepackage{amssymb,amsmath,amsthm}
\usepackage{mathrsfs}
\DeclareMathAlphabet\EuScript{U}{eus}{m}{n}
\SetMathAlphabet\EuScript{bold}{U}{eus}{b}{n}

\newcommand{\ie}{\emph{i.e.}}
\newcommand{\eg}{\emph{e.g.}}
\newcommand{\cf}{\emph{cf.}}
\newcommand{\etal}{\emph{et al.}}

\newcommand{\viz}{\emph{viz.}}
\newcommand{\B}{\mathscr{B}}
\newcommand{\K}{\mathcal{K}}
\newcommand{\C}{\mathbb{C}}
\def\H{\mathcal H}
\newcommand{\N}{\mathbb{N}}
\newcommand{\R}{\mathbb{R}}
\newcommand{\e}{\mathrm{e}}
\newcommand{\de}{\mathrm{d}}

\newcommand{\strip}{\Gamma}

\newcommand{\sii}{L^2}
\newcommand{\sinf}{L^\infty}
\newcommand{\Dom}{\matheucal D}

\newcommand{\supp}{\mathop{\mathrm{supp}}\nolimits}

\newcommand{\diag}{\mathop{\mathrm{diag}}\nolimits}
\newcommand{\tr}{\mathop{\mathrm{tr}}\nolimits}
\newcommand{\esssup}{\mathop{\mathrm{ess\,sup}}}
\newcommand{\threshold}{\mathcal{T}}
\newcommand{\identity}{1}
\newtheorem{Theorem}{Theorem}[section]
\newtheorem{Proposition}[Theorem]{Proposition}
\newtheorem{Lemma}[Theorem]{Lemma}
\newtheorem{Corollary}[Theorem]{Corollary}
\newtheorem{Remark}[Theorem]{Remark}
\newtheorem{Definition}[Theorem]{Definition}
\newtheorem{ass}{Assumption}[section]
\newenvironment{Assumption}{\begin{ass}}{\end{ass}}
\begin{document}
%
%
\title{{\Large\textbf{
The nature of the essential spectrum in curved \\ quantum waveguides
\footnote{Published in J.\ Phys.~A \textbf{37} (2004), no.~20, pp.~5449--5466.}
}}
}
\author{David Krej\v{c}i\v{r}\'{\i}k$\,^{1}$%
\footnote{
On leave of absence from
\emph{Department of Theoretical Physics,
Nuclear Physics Institute, Academy of Sciences,
250\,68 \v{R}e\v{z} near Prague, Czech Republic}.
} 
\ \ and \ Rafael Tiedra de Aldecoa$\,^{2}$}
\date{\small 
\begin{quote}
\emph{
\begin{itemize}
\item[$^1$]
Departamento de Matem\'atica, Instituto Superior T\'ecnico, \\
Av. Rovisco Pais, 1049-001 Lisboa, Portugal
\item[$^2$]
D\'epartement de Physique Th\'eorique, Universit\'e de Gen\`eve, \\
24, quai E. Ansermet, 1211 Gen\`eve 4, Switzerland
\item[]
\emph{E-mail\;\!:} dkrej@math.ist.utl.pt \,and\, rafael.tiedra@physics.unige.ch
\end{itemize}
}
\end{quote}
28 March 2004
}
\maketitle
%
%
\begin{abstract}
\noindent
We study the nature of the essential spectrum of the Dirichlet Laplacian
in tubes about infinite curves embedded in Euclidean spaces.
Under suitable assumptions about the decay of curvatures at infinity,
we prove the absence of singular continuous spectrum
and state properties of possible embedded eigenvalues.
The argument is based on Mourre conjugate operator method
developed for acoustic multistratified domains
by Benbernou in~\cite{Benbernou2} and Dermenjian \etal\/ in~\cite{DDI}.
As a technical preliminary, we carry out a spectral analysis
for Schr\"odinger-type operators in straight Dirichlet tubes.
We also apply the result to the strips embedded in abstract surfaces.
\end{abstract}
\newpage
%
%
\setcounter{equation}{0}
\section{Introduction}
A strong physical motivation to study the Dirichlet Laplacian 
in infinitely stretched tubular regions
comes from the fact it constitutes a reasonable model 
for the Hamiltonian of a non-relativistic quantum particle 
in mesoscopic systems called \emph{quantum waveguides}~\cite{DE,LCM,Hurt}.
Since there exists a close relation between spectral 
and scattering properties of Hamiltonians,
one is naturally interested in carrying out the spectral analysis 
of the Laplacian in order to understand the quantum dynamics in waveguides.
For instance, the crucial step in most proofs of asymptotic completeness
is to show that the Hamiltonian has no singular continuous spectrum~\cite{RS3}.
The Laplacian in a tube has attracted considerable attention
since it was shown
in \cite{ES} that there may be discrete eigenvalues in curved waveguides.
However, a detailed analysis of the essential part of the spectrum
has been left aside up to now.
The purpose of the present paper is to fill in this gap.
 
The usual model for a curved quantum waveguide,
which we adopt in this paper, is as follows.
Let~$s \mapsto p(s)$ be an infinite unit-speed smooth curve in~$\R^d$, 
$d \geq 2$ (the physical cases corresponding to~$d=2,3$).
Assuming that the curve possesses an appropriate 
smooth Frenet frame $\{e_1,\dots,e_d\}$
(\cf~Assumption~\ref{Ass.Frenet}),
the $i^\mathrm{th}$ curvature~$\kappa_i$ of~$p$, $i\in\{1,\dots,d-1\}$,  
is a smooth function of the arc-length parameter $s\in\R$. 
Given a bounded open connected set~$\omega$ in~$\R^{d-1}$
with the centre of mass at the origin, we identify the configuration
space~$\strip$ of the waveguide with a tube of
cross-section~$\omega$ about~$p$, namely\:\!:
\begin{equation}\label{tube}
  \Gamma := \mathscr{L}(\R\times\omega)\:\!,
  \qquad
  \mathscr{L}(s,u^2,\dots,u^d) 
  := p(s)+u^\mu \:\! \mathcal{R}_\mu^{\ \nu}(s) \:\! e_\nu(s)\:\!,
\end{equation}
where $\mu,\nu$ are summation indices taking values in $\{2,\ldots,d\}$
and $(\mathcal{R}_\mu^{\ \nu})$ is a family of rotation matrices in~$\R^{d-1}$.
In this paper, we choose the rotations in such a way
that $(s,u)$, with $u:=(u^2,\dots,u^d)$, 
are orthogonal ``coordinates" (\cf~Section~\ref{SubSec.tubes}) 
due to the technical simplicity.
It should be stressed here that while the shape of the tube~$\strip$ is
not influenced by a special choice of~$(\mathcal{R}_\mu^{\ \nu})$ 
provided~$\omega$ is circular, this may no longer be true for a general
cross-section. We make the hypotheses (Assumption~\ref{Ass.basic})
that~$\kappa_1$ is bounded, 
$
  a \|\kappa_1\|_\infty < 1 ,
$
with $a:=\sup_{u\in\omega}|u|$,
and~$\strip$ does not overlap itself
so that the tube can be globally parameterised by~$(s,u)$.
Our object of interest is the Dirichlet Laplacian
associated with the tube, \ie,
\begin{equation}\label{laplacian}
  -\Delta_{\rm D}^{\strip}
  \qquad\textrm{on}\qquad
  \sii(\strip).
\end{equation}

If~$p$ is a straight line, \ie,~all~$\kappa_i = 0$,
then~$\strip$ may be identified
with the straight tube $\Omega:=\R\times\omega$.
In that case, it is easy to see that the spectrum of~(\ref{laplacian})
is purely \emph{absolutely continuous} and equal to the
interval~$[\nu_1,\infty)$, where~$\nu_1$ denotes the first eigenvalue
of the Dirichlet Laplacian in the cross-section~$\omega$.

On the other hand, if~$p$ is non-trivially curved
and straight asymptotically,
in the sense that the curvature~$\kappa_1$
vanishes at infinity, then the essential spectrum of~(\ref{laplacian})
remains equal to~$[\nu_1,\infty)$. However,
there are always \emph{discrete eigenvalues} below~$\nu_1$.
When \mbox{$d=2$}, the latter was proved for the first time in~\cite{ES}
for a rapidly decaying curvature and sufficiently small~$a$.
Numerous subsequent studies improved
and generalised this initial result \cite{GJ,RB,DE,K1,KKriz,CDFK}.
The generalisation to tubes of circular cross-section in~$\R^3$
was done in~\cite{GJ} (see also~\cite{DE})
and the case of any dimension~$d \geq 2$
and arbitrary cross-section can be found in~\cite{CDFK}.
Let us also mention that the discrete spectrum may be generated by
other local perturbations of the straight tube~$\Omega$
(see, \eg, \cite{BGRS,EV2,BEGK}), but in the bent-tube case the phenomenon
is of a purely quantum origin because there are no classical closed
trajectories, apart from those given by a zero measure set of initial
conditions in the phase space.

The main goal of the present work is a thorough analysis
of the \emph{essential spectrum} of~(\ref{laplacian}).
In particular, we find sufficient conditions
which guarantee that the essential spectrum of a curved tube
``does not differ too much'' from the straight case
(for simplicity, we present here our results only for~$d=2$,
see Theorem~\ref{Thm.tubes} for the~$d$-dimensional case)\:\!:
\begin{Theorem}[$\boldsymbol d$=2]\label{Thm}
Let~$\strip$ be as above for $d=2$ ($\kappa := \kappa_1$)
and~$\threshold:=\{n^2\nu_1\}_{n=1}^\infty$ with~$\nu_1:=\pi^2/(2a)^2$ 
(the set of eigenvalues of the Dirichlet Laplacian
in the~$1$-di\-men\-sional cross-section~$\omega$). 
Suppose
\begin{enumerate}
\item $\kappa(s)$,~$\ddot{\kappa}(s)$ 
$\longrightarrow0$~~as~~$|s|\to\infty$\:\!,
\item $\exists\:\!\vartheta\in(0,1]$~~s.t.~~$\dot{\kappa}(s)$,~$\dddot{\kappa}(s)$
$=\mathcal{O}\left(|s|^{-(1+\vartheta)}\right)$\:\!.
\end{enumerate}
Then
\begin{itemize}
\item[\emph{(i)}]
$
  \sigma_\mathrm{ess}\left(-\Delta_{\rm D}^{\strip}\right)
  =[\nu_1,\infty)
$\:\!,
  \item[\emph{(ii)}]
  $\sigma_\mathrm{sc}\left(-\Delta_{\rm D}^{\strip}\right)=\varnothing$\:\!,
\item[\emph{(iii)}]
  $\sigma_\mathrm{p}\left(-\Delta_{\rm D}^{\strip}\right)
  \cup \threshold$~is closed and countable\:\!,
\item[\emph{(iv)}]
$\sigma_\mathrm{p}\left(-\Delta_{\rm D}^{\strip}\right)
  \setminus \threshold$
  is composed of finitely degenerated eigenvalues which can accumulate at points
of~$\threshold$ only\:\!.
\end{itemize}
\end{Theorem}
\noindent
To prove this theorem (and the general Theorem~\ref{Thm.tubes}), 
we use the conjugate operator method
introduced by~\cite{Mourre} E.~Mourre
and lastly developed by~\cite{ABG} W.~Amrein \etal\/ Notice that
the set~$\threshold$ plays a role analogous to the set of \emph{thresholds}
in the Mourre theory of~$N$-body Schr\"odinger operators~\cite{CFKS}. 

Actually, the property~(i) holds true whenever
the first curvature vanishes at infinity,
without assuming any decay of the derivatives  
(they may not even exist),
see~\cite{KKriz} for $d=2$ and~\cite{CDFK} for the general case.
Our second result~(ii) can be compared only with~\cite{DES} (see also~\cite{DEM}),
where the problem of resonances is investigated for~$d=2$.
Assuming that there exists~$\vartheta\in(0,1]$ such that
$
  \kappa(s), \dot{\kappa}(s)^2, \ddot{\kappa}(s)
  =\mathcal{O}\big(|s|^{-(1+\vartheta)}\big)
$,  
the authors proved the absence of singular continuous spectrum
as a consequence of the completeness of wave operators
obtained by standard smooth perturbation methods of scattering theory.
Notice that our and their results are independent. Indeed,
while we need to require a faster decay of~$\dot\kappa$  
and also impose a condition on~$\dddot\kappa$,
our decay assumptions on~$\kappa$ and~$\ddot\kappa$ 
are on the contrary much weaker.
Our other spectral results~(iii) and~(iv) 
(and~(ii) for~$d \geq 3$) are new.

The organisation of this paper is as follows.
In Section~\ref{Sec.straight},
we consider the Schr\"o\-ding\-er-type operator
\begin{equation}\label{hamiltonian}
  H := -\partial_i\:\!G^{ij}\partial_j + V
  \qquad\textrm{on}\qquad
  \H(\Omega):=\sii(\Omega)\:\!,
\end{equation}
subject to Dirichlet boundary conditions,~$i$ and $j$ being summation
indices taking values in $\{1,\ldots,d\}$,~$G\equiv(G^{ij})$ a real
symmetric matrix-valued measurable function on~$\Omega$ and~$V$
the multiplication operator by a real-valued measurable
function on~$\Omega$.
We make Assumption~\ref{Ass1} and Assumption~\ref{Ass2} stated below.
Adapting the approach of~\cite{Benbernou2,DDI}
to non-zero~$V$ and~$G$ different from a multiple of the identity,
we study the nature of the essential spectrum of the operator~$H$.
In particular, we prove the absence 
of singular continuous spectrum and state properties 
of possible embedded eigenvalues. 
The result is contained in Theorem~\ref{Thm.main}
and is of independent interest.
In Section~\ref{Sec.tubes},
we apply it to the case of curved tubes~(\ref{tube}).
Using the diffeomorphism 
$
  \mathscr{L}: \Omega \to \strip
$
and a unitary transformation
(ideas which go back to~\cite{ES}), we cast the
Laplacian~(\ref{laplacian}) into a unitarily equivalent
operator of the form~(\ref{hamiltonian})
for which Theorem~\ref{Thm.main} can be used.
The obtained spectral results can be found in Theorem~\ref{Thm.tubes}
(the general version of Theorem~\ref{Thm} above).
Finally, in Section~\ref{Sec.strips},
we similarly investigate the essential spectrum
of the Dirichlet Laplacian in an infinite strip
in an abstract two-dimensional Riemannian manifold of curvature~$K$.
The general result is contained in Theorem~\ref{Thm.general},
while the case of flat strips, \ie,~with~$K = 0$,
is summarised in Theorem~\ref{Thm.flat}
(the latter involves the curved strips in~$\R^2$
as a special case).

For the conjugate operator method and notation
used in Section~\ref{Sec.straight},
the reader is referred to~\cite{ABG}
and particularly to short well-arranged reviews of the abstract theory
in~\cite[Sec.~2]{Benbernou2} or~\cite[Sec.~1]{DDI}.
A more detailed geometric background for Section~\ref{Sec.tubes}
and Section~\ref{Sec.strips} can be found in~\cite{Kli,CDFK}
and~\cite{Gray,K1}, respectively.

We use the standard component notation of tensor analysis
throughout the paper.
In particular, the repeated indices convention is adopted henceforth,
the range of indices being~$1, \dots, d$~for Latin and~$2,\dots,d$~for Greek.
The indices are associated in a natural way with the components
of~$x\in\R\times\omega$.
The partial derivative w.r.t.~$x^i$ is often denoted 
by a comma with the index~$i$.
The brackets~$(\cdot)$ are used in order to distinguish a matrix
from its coefficients. The symbols~$\delta_{ij}$ and~$\delta^{ij}$  
are reserved for the components of the identity matrix~$1$.

\setcounter{equation}{0}
\section{Schr\"odinger-type operators in straight tubes}\label{Sec.straight}
\subsection{Preliminaries}\label{Sec.Preliminaries}

Let~$\omega$ be an (arbitrary) bounded open connected
set in~$\R^{d-1}$,~$d \geq 2$, and consider the
straight tube~$\Omega:=\R\times\omega$.
Our object of interest in this section is the operator
given formally by~(\ref{hamiltonian}), subject to
Dirichlet boundary conditions. In addition to the basic
properties required for the matrix~$G$ and function~$V$,
we make the following assumptions.
\begin{Assumption}\label{Ass1}
\rule{0.5\textwidth}{0ex}
\begin{enumerate}
\item $\exists\:\!C_\pm\in(0,\infty)$~~s.t.~~$C_-
\:\!\identity \leq G(x) \leq C_+\:\!\identity$~~for
a.e.~$x\in\Omega$\:\!,
\item 
$\forall i,j\in\{1,\dots,d\}$\:\!,~~${\displaystyle\lim_{R\to\infty}} \
{\displaystyle\esssup_{x\in\left(\R\setminus[-R,R]\right)\times\omega}}
\left|G^{ij}(x)-\delta^{ij}\right|=0$\:\!,
\item $\exists\:\!\vartheta_1\in(0,1]$\:\!,~$C\in(0,\infty)$~~s.t.~$\big(\big|G^{ij}_{\ ,1}(x)
\big|\big)\leq C\:\!\langle x^1\rangle^{-(1+\vartheta_1)}\:\!\identity$~~for a.e.~$x\in\Omega$\:\!,
\item
$G^{1i}_{\ ,i}\in L^\infty(\Omega)$\:\!.
\end{enumerate}
\end{Assumption}
\noindent
Here~$\langle\cdot\rangle:=\left(1+|\cdot|^2\right)^{1/2}$
and the inequalities must be understood in the sense
of matrices.
\begin{Assumption}\label{Ass2}
\rule{0.5\textwidth}{0ex}
\begin{enumerate}
\item $V\in\sinf(\Omega)$\:\!,
\item 
$
{\displaystyle\lim_{R\to\infty}} \
{\displaystyle\esssup_{x\in\left(\R\setminus[-R,R]\right)\times\omega}} |V(x)| = 0
$\:\!,
\item $\exists\:\!\vartheta_2\in(0,1]$\:\!,~$C\in(0,\infty)$~~s.t.~~$\left|V_{,1}(x)\right|
\leq C\:\!\langle x^1 \rangle^{-(1+\vartheta_2)}$~~for a.e.~$x\in\Omega$\:\!.
\end{enumerate}
\end{Assumption}

Let us fix some notations. 
We write~$\H^\nu(\Omega)$ and~$\H_0^\nu(\Omega)$,
$\nu\in\R$, for the usual Sobolev spaces
\cite{Ad}. Given two Hilbert spaces~$\H_1$ and~$\H_2$, we
denote by~$\B(\H_1,\H_2)$,
respectively~$\K(\H_1,\H_2)$, the set of bounded,
respectively compact, operators from~$\H_1$ to~$\H_2$.
We also define~$\B(\H_1):=\B(\H_1,\H_1)$
and~$\K(\H_1):=\K(\H_1,\H_1)$.
We denote by $\H_1^*$ the topological
antidual of $\H_1$. We write~$(\cdot,\cdot)$
for the inner product in $\H(\Omega)$
and~$\|\cdot\|$ for the norm in~$\H(\Omega)$ and
$\B\left(\H(\Omega)\right)$.

We now give a meaning to the formal expression~(\ref{hamiltonian}).
We start by introducing the sesquilinear form~$Q_0$ on~$\H(\Omega)$
defined by
\begin{equation}
  Q_0(\varphi,\psi):=\left(\varphi_{,i},\delta^{ij}\psi_{,j}\right),
  \qquad
  \varphi,\psi\in\Dom(Q_0):=\H_0^1(\Omega)\:\!,
\end{equation}
which is densely defined, symmetric, non-negative and closed. Consequently,
there exists a unique self-adjoint operator~$H_0$
associated with it, which is just the Dirichlet
Laplacian~$-\Delta_{\rm D}^\Omega$ on~$\sii(\Omega)$. We have
\begin{equation*}
  H_0\psi=-\Delta\psi\:\!,
  \qquad
  \psi\in\Dom(H_0)=
  \left\{\psi\in\H_0^1(\Omega)\,:\,
  \Delta\psi\in\H(\Omega)\right\}.
\end{equation*}
We consider~$H$ as an operator obtained
by perturbing the free Hamiltonian~$H_0$.
Since the matrix~$G$ is uniformly positive and bounded
by Assumption~\ref{Ass1}.1, the sesquilinear form
$(\varphi,\psi)\mapsto\left(\varphi_{,i},G^{ij}\psi_{,j}\right)$
defined on~$\Dom(Q_0)\times\Dom(Q_0)$ is also densely defined,
symmetric, non-negative and closed. At the same time, the
potential~$V$ is supposed to be bounded by Assumption~\ref{Ass2}.1,
which means that the sesquilinear form~$Q$ defined by
\begin{equation}
  Q(\varphi,\psi):=\left(\varphi_{,i},G^{ij}\psi_{,j}\right)
  +\left(\varphi,V\psi\right),
  \qquad
  \varphi,\psi\in\Dom(Q):=\H_0^1(\Omega)\:\!,
\end{equation}
gives rise to a semi-bounded self-adjoint operator~$H$.
Using the representation theorem \cite[Chap.~VI, Thm.~2.1]{Kato}
and the fact that~$V$ is bounded (recall also Assumption~\ref{Ass1}.1), 
one may check that
\begin{equation*}
  \Dom(H)=
  \left\{\psi\in\H_0^1(\Omega)\,:\,
  \partial_i\:\!G^{ij}\:\!\partial_j \psi\in\H(\Omega) \right\},
\end{equation*}
where the derivatives must be interpreted in the distributional sense,
and that $H$ is acting as in~(\ref{hamiltonian}) on its domain.

For any~$z\in\C\setminus\sigma(H_0)$~, respectively~$z\in\C\setminus\sigma(H)$,
let~$R_0(z):=(H_0-z)^{-1}$, respectively~$R(z):=(H-z)^{-1}$. 

\subsection{Localisation of the essential spectrum}\label{localisation}

The Dirichlet Laplacian~$-\Delta_{\rm D}^\omega$ on~$\sii(\omega)$,
\ie,~the operator associated with
\begin{equation*}
  q(\varphi,\psi):=\left(\varphi_{,\mu},\delta^{\mu\nu}\psi_{,\nu}\right),
  \qquad
  \varphi,\psi\in\Dom(q):=\H_0^1(\omega)\:\!,
\end{equation*}
has a purely discrete spectrum consisting of 
eigenvalues $\nu_1<\nu_2\leq\nu_3\leq\dots$ with $\nu_1 > 0$.
We set $\threshold:=\{\nu_n\}_{n=1}^\infty$.
Since~$H_0$ is naturally decoupled in the following way\:\!:
\begin{equation*}
  H_0=-\Delta^\R \otimes\identity + \identity\otimes (-\Delta_{\rm D}^\omega)
  \qquad\textrm{on}\qquad\sii(\R) \otimes \sii(\omega)\:\!,
\end{equation*}
where~``$\otimes$'' denote the closed tensor product,~$\identity$ the identity
operators on appropriate spaces and~$-\Delta^\R$ the
Laplacian on~$\sii(\R)$, one has
\begin{equation}\label{EssSp0}
  \sigma(H_0)=\sigma_\mathrm{ess}(H_0)=[\nu_1,\infty)\:\!.
\end{equation}
In order to prove that (under our assumptions) $H$ possesses
the same essential spectrum, we need the following lemma.
\begin{Lemma}\label{phi compact}
  Let $\varphi\in C^\infty_0(\R)$ and set
  $\phi:=\varphi\otimes1$ on $\Omega$.
  Then, as a multiplication operator,
  $\phi\in\K\left(\Dom(H_0),\H^1_0(\Omega)\right)$. 
\end{Lemma}
\begin{proof}
  Since 
  \begin{equation*}
    \phi=H_0^{-1/2}\:\!H_0^{1/2}\:\!\phi\:\!H_0^{-1}H_0
  \end{equation*}
  in $\B\left(\Dom(H_0),\H^1_0(\Omega)\right)$,
  $H_0\in\B\left(\Dom(H_0),\H(\Omega)\right)$ and
  $H_0^{-1/2}\in\B\left(\H(\Omega),\H^1_0(\Omega)\right)$,
  it is enough to prove that
  $H_0^{1/2}\phi\:\!H_0^{-1}\in\K\left(\H(\Omega)\right)$.
  However,
  \begin{align}
    H_0^{1/2}\:\!\phi\:\!H_0^{-1}
    &=H_0^{-1/2}\left[H_0,\phi\right]H_0^{-1}+H_0^{-1/2}\:\!\phi\nonumber\\
    &=-H_0^{-1/2}\left(2\:\!\phi_{,1}\:\!\partial_1+\phi_{,11}\right)H_0^{-1}
    +H_0^{-1/2}\:\!\phi\:\!,\label{compact terms}
  \end{align}
  where each term on the r.h.s. is in $\K\left(\H(\Omega)\right)$.
  Let us demonstrate it for the first term.
  Since $\partial_1H_0^{-1}\in\B\left(\H(\Omega)\right)$, it is
  sufficient to prove that
  $H_0^{-1/2}\phi_{,1}\in\K\left(\H(\Omega)\right)$.
  Let $z_1\in(-\infty,0)$ and $z_2\in(-\infty,\nu_1)$
  be such that $z_1+z_2=0$. Define
  $R_\shortparallel(z_1):=\left(-\Delta^{\mathbb R}-z_1\right)^{-1}$
  and $R_\perp(z_2):=\left(-\Delta^\omega_{\mathrm D}-z_2\right)^{-1}$.
  Then, using some standard results on tensor
  products of operators \cite[Chap.~11]{KR}, one can write
  \begin{equation*}
    H_0^{-1/2}\phi_{,1}
    =H_0^{-1/2}\big[R_\shortparallel^{-1/4}(z_1)
      \otimes R_\perp^{-1/4}(z_2)\big]
    \big[R_\shortparallel^{1/4}(z_1)\:\!\varphi_{,1}
      \otimes R_\perp^{1/4}(z_2)\big]
  \end{equation*}
  where~$\varphi_{,1}$ is viewed
  as a multiplication operator
  in~$\sii(\R)$. The third factor
  on the r.h.s.~is in $\K\left(\H(\Omega)\right)$
  because $-\Delta^\omega_{\mathrm D}$ has a compact resolvent
  and $R_\shortparallel^{1/4}(z_1)\:\!\varphi_{,1}
  \in\mathcal K\left(\sii(\mathbb R)\right)$
  by~\cite[Thm.~4.1.3]{ABG}. 
  The remaining factors can be rewritten as
  \begin{equation*}
    \Psi(X_1,X_2):=(X_1+X_2)^{-1/2}X_1^{1/4}X_2^{1/4}
  \end{equation*}
  with~$X_1:=\left(-\Delta^{\mathbb R}-z_1\right)\otimes\identity$
  and~$X_2:=\identity\otimes\left(-\Delta^\omega_{\mathrm D}-z_2\right)$ (both self-adjoint and
  mutually commuting). So, one can estimate
  \begin{equation*}
    \left\|\Psi(X_1,X_2)\right\|
    \leq\sup_{x_1,x_2\in(0,\infty)}(x_1+x_2)^{-1/2}(x_1x_2)^{1/4}<\infty\:\!.
  \end{equation*}
  Hence, the first term on the r.h.s. of~(\ref{compact terms})
  is in $\K\left(\H(\Omega)\right)$. The argument is similar
  for the remaining terms.
\end{proof}
\begin{Proposition}\label{Prop.compact} One has
  \begin{itemize}
  \item[\emph{(i)}] $\forall z\in\C\setminus\left(\sigma(H)\cup
    \sigma(H_0)\right)$\:\!,~~$R(z)-R_0(z)\in\K\big(\H(\Omega)\big)$\:\!,
  \item[\emph{(ii)}] $\sigma_\mathrm{ess}(H)=[\nu_1,\infty)$\:\!.
  \end{itemize}
\end{Proposition}
\begin{proof}
  We prove (i) for some (and hence for all) value 
  of~$z\in\C\setminus\left(\sigma(H)\cup
  \sigma(H_0)\right)$. Let
  $z\in\C\setminus\R$. Define $R_1(z):=\left(H_0+V-z\right)^{-1}$.
  Then, one has
  \begin{equation*}
    R(z)-R_0(z)=R(z)-R_1(z)-R_1(z)\:\!VR_0(z)\:\!.
  \end{equation*}
  Let us first consider ${R(z)-R_1(z)}$. Knowing that
  $H$ and ${H_0+V}$ have the same form domain, the identity
  \begin{equation*}
    R(z)-R_1(z)=-R(z)\:\!(H-H_0-V)\:\!R_1(z)
  \end{equation*}
  holds in $\B\left(\H^{-1}(\Omega),\H^1_0(\Omega)\right)$. But,
  one has the following sequence of continuous and dense
  imbeddings of Hilbert spaces
  \begin{equation*}
    \Dom(H)\subset\H^1_0(\Omega)\subset\H(\Omega)
    \subset\H^{-1}(\Omega)\subset\Dom(H)^*
  \end{equation*}
  which implies that $R(z)$ extends (by duality) to a
  homeomorphism of $\Dom(H)^*$ onto $\H(\Omega)$. Thus,
  since $R_1(z)$ is also a homeomorphism from $\H(\Omega)$
  onto $\Dom(H_0)$,
  ${R(z)-R_1(z)}\in\K\left(\H(\Omega)\right)$
  if and only if ${H-H_0-V}\in\K\left(\Dom(H_0),\Dom(H)^*\right)$.
  For all $n\in\N\setminus\{0\}$, let
  $\varphi_n\in C^\infty_0(\R)$ be such that
  $0\leq\varphi_n\leq1$ and
  \begin{equation*}
    \varphi_n(x^1)=
    \begin{cases}
      1 & \textrm{if} \quad |x^1| \leq n\\
      0 & \textrm{if} \quad |x^1| \geq n+1\:\!.
    \end{cases}  
  \end{equation*}
  Set $\phi_n:=\varphi_n\otimes\identity$ on $\Omega$ and
  \begin{equation*}
    K_n\:\!\psi:=-\partial_i\:\!F^{ij}\:\!\phi_n\:\!\partial_j\:\!\psi\:\!,
    \qquad\psi\in\Dom(H_0)\:\!,
  \end{equation*}  
  where~$(F^{ij}):=(G^{ij}-\delta^{ij})$. Clearly, ${H-H_0-V},
  K_n\in\B\left(\Dom(H_0),\Dom(H)^*\right)$ and 
  \begin{align*} 
    &\left\|K_n-(H-H_0-V)\right\|_{\B\left(\Dom(H_0),\Dom(H)^*\right)}
    \\
    &\equiv\sup_{\psi\in\Dom(H_0),\:\!\|\psi\|_{\Dom(H_0)}=1}
    \big\|\left(1+H^2\right)^{-1/2}
    \left[-\partial_i\:\!F^{ij}\:\!(\phi_n-1)\:\!\partial_j\right]\psi\big\|
    \\
    &\leq\sup_{\psi\in\Dom(H_0),\:\!\|\psi\|_{\Dom(H_0)}=1}
    \sum_{j=1}^d
    \big\|\left(1+H^2\right)^{-1/2}\partial_i\big\|
    \left\|F^{ij}\:\!(\phi_n-1)\right\|_\infty
    \left\|\psi\right\|_{\H^1_0(\Omega)}
    \xrightarrow[n\to\infty]{}0\:\!,
  \end{align*}
  where we have used the fact that $\Dom(H_0)\subset\H^1_0(\Omega)$
  continuously and Assumption~\ref{Ass1}.2 in the final step.
  So, it only remains to show that
  $K_n\in\K\left(\Dom(H_0),\Dom(H)^*\right)$. After a commutation,
  one gets in $\B\left(\Dom(H_0),\Dom(H)^*\right)$
  \begin{equation*}
    K_n=-\partial_i\:\!F^{ij}\:\!\partial_j\:\!\phi_n
    +\partial_i\:\!F^{i1}\:\!\phi_{n,1}
  \end{equation*}
  where $\phi_n$, $\phi_{n,1}$ are seen as multiplication operators
  in $\H(\Omega)$. It is clear that both
  $\partial_i\:\!F^{ij}\partial_j$ and $\partial_i\:\!F^{i1}$
  are in $\B\left(\H^1_0(\Omega),\Dom(H)^*\right)$.
  Moreover, $\phi_n$ and $\phi_{n,1}$ are in
  $\K\left(\Dom(H_0),\H^1_0(\Omega)\right)$ by Lemma~\ref{phi compact}.
  Thus, $K_n\in\K\left(\Dom(H_0),\Dom(H)^*\right)$ so that
  ${R(z)-R_1(z)}\in\K\left(\H(\Omega)\right)$. Using similar arguments,
  one can also prove that the $R_1(z)\:\!V\phi_n\:\!R_0(z)$ is compact
  and converges to $R_1(z)\:\!VR_0(z)$ in $\B(\H(\Omega))$ due to
  Assumption~\ref{Ass2}.2. This implies that
  $R_1(z)\:\!VR_0(z)\in\K\left(\H(\Omega)\right)$.
  
  \noindent
  (ii) It is a direct consequence of (i), (\ref{EssSp0})
  and Weyl's theorem~\cite[Thm.~XIII.14]{RS4}.
\end{proof}
\begin{Remark}
  Notice that \emph{Assumptions~\ref{Ass1}.3},~\emph{\ref{Ass1}.4}
  and~\emph{\ref{Ass2}.3} are not used in the proof of
  \emph{Proposition~\ref{Prop.compact}}.
\end{Remark}
%
\subsection{Nature of the essential spectrum}\label{Sec.nature}
%
This part is devoted to a more detailed analysis of the essential
spectrum of~$H$. In particular, we show that the singular
continuous spectrum is empty.
The strategy adapted from~\cite{Benbernou2} is the following.
Firstly, we construct a dilation operator~$A$
such that~$H_0\in C^\infty(A)$ and~$H\in C^{1+\vartheta}(A)$
with~$\vartheta:= \min\{\vartheta_1,\vartheta_2\}\in(0,1]$
(see~\cite{ABG}, \cite[Sec.~2]{Benbernou2} or~\cite[Sec.~1]{DDI}
for definitions of the spaces involved here and in the
sequel).
Secondly, we prove that~$A$ is \emph{strictly conjugate}
(in Mourre's sense) to~$H_0$ on~$\R\setminus\threshold$.
Finally, since~$R(i)-R_0(i)$ is compact by 
the first claim of Proposition~\ref{Prop.compact}
and both~$H$ and~$H_0$ are of class
$
  C_u^1(A)
  \supseteq C^{1+\vartheta}(A)
  \supseteq C^\infty(A)
$,
it follows that~$A$ is conjugate to~$H$
on~$\R\setminus\threshold$ as well.

\subsubsection{The dilation operator}

Let~$q^1$ be the multiplication operator by the coordinate~$x^1$
in~$\H(\Omega)$. Let
\begin{equation}\label{dilation}
  A:=\mbox{$\frac12$} \left(q^1 p_1 + p_1q^1\right)
  \qquad\textrm{with}\qquad
  p_1:=-i\:\!\partial_1
\end{equation}
be the dilation operator in~$\H(\Omega)$
w.r.t.~$x^1$, \ie,~the self-adjoint extension of the operator
defined by the expression~(\ref{dilation})
with~$C_0^\infty(\Omega)$ as initial domain. Define
$A_\shortparallel$ as the self-adjoint operator in $\sii(\R)$ such that
$A=A_\shortparallel\otimes 1$.
\begin{Remark}\label{Rem.isomorphism}
  The group~$\left\{\e^{iAt}\right\}_{t\in\R}$ leaves
  invariant~$\H^1_0(\Omega)$. Indeed, using the natural
  isomorphism~$\H^1_0(\Omega)\simeq\H^1(\R)\otimes\H^1_0(\omega)$,
  one can write
\begin{equation*}
  \forall t\in\R\:\!,\qquad
  \e^{iAt}\;\!\H^1_0(\Omega)
  = \big(\e^{iA_\shortparallel t}\:\!\H^1(\R)\big)
  \otimes\H^1_0(\omega)\:\!.
\end{equation*}
Then, the affirmation follows from the fact
\emph{\cite[Prop.~4.2.4]{ABG}} that~$\H^1(\R)$ is  stable
under~$\big\{\e^{iA_\shortparallel t}\big\}_{t\in\R}$.
\end{Remark}
In order to deal with the commutator~$i\left[H,A\right]$,
we need the following family of operators 
\begin{equation}\label{regular}
  \left\{p_1(\varepsilon)  
  := p_1(1+i\varepsilon p_1)^{-1}\right\}_{\varepsilon>0}\:\!,
\end{equation}
which regularises the momentum operator~$p_1$\:\!:
\begin{Lemma}\label{lem.regular}
  One has
  \begin{itemize}
  \item[\emph{(i)}]
    $
    \left\{p_1(\varepsilon)\right\}_{\varepsilon>0}\subset\B\big(\H(\Omega)\big)
    $\:\!,
  \item[\emph{(ii)}]
    $\left\{p_1(\varepsilon)\right\}_{\varepsilon>0}$ is uniformly
    bounded in $\B\big(\H^1(\Omega),\H(\Omega)\big)$

    and \emph{s}-$\displaystyle\lim_{\varepsilon\to0}p_1
    (\varepsilon)=p_1$ in $\B\big(\H^1(\Omega),\H(\Omega)\big)$\:\!,
  \item[\emph{(iii)}]~$\forall\varepsilon>0$\:\!,~~$\left[p_1(\varepsilon),q_1\right]
    =-i\:\!(1+i\varepsilon p_1)^{-2}$ in $\B(\H(\Omega))$\:\!,
  \item[\emph{(iv)}]~$\forall\varepsilon>0$\:\!,~~$p_1(\varepsilon)\:\!
    \H_0^1(\Omega) \subset \H_0^1(\Omega)$\:\!. 
  \end{itemize}
\end{Lemma}
\begin{proof}
The first three assertions are established in \cite[Lemma~4.1]{Benbernou2}.
Consequently, it only remains to prove the last statement.
Using the isomorphism mentioned in Remark~\ref{Rem.isomorphism},
one can write
\begin{equation*}
  \forall\varepsilon>0\:\!,\qquad p_1(\varepsilon)\H^1_0(\Omega)
  =-i\varepsilon^{-1}\big\{\big[1+i\varepsilon^{-1}
    (p_1-i\varepsilon^{-1})^{-1}\big]\H^1(\R)\big\}\otimes\H^1_0(\omega)\:\!,
\end{equation*}
where~$p_1$ on the r.h.s. must be viewed as an operator acting
in~$\sii(\R)$.  With this last relation, it is clear
that~$\H_0^1(\Omega)$ is left invariant by the
family~$\left\{p_1(\varepsilon)\right\}_{\varepsilon>0}$\:\!.
\end{proof}
We also need the following density result for the set
$\Dom(H)_c:=\{\psi\in\Dom(H)\,:\,\supp(\psi)\textrm{ is compact}\:\!\}$.
\begin{Lemma}\label{density}
  One has
  \begin{itemize}
    \item[\emph{(i)}] $\Dom(H)_c$ is dense in $\Dom(H)$\:\!,
    \item[\emph{(ii)}] $\Dom(H)_c$ is dense in $\H^1_0(\Omega)$\:\!.
  \end{itemize}
\end{Lemma}
\begin{proof}
  (i) We are inspired by \cite[Lemma~2.1]{DDI}.
  Let $\psi\in\Dom(H)$. Define $\varphi_0\in C^\infty_0(\R)$
  such that 
  \begin{equation*}
    \varphi_0(x^1)=
    \begin{cases}
      1 & \textrm{if} \quad |x^1| \leq 1\\
      0 & \textrm{if} \quad |x^1| \geq 2\:\!.
    \end{cases}  
  \end{equation*}
  Let $n\in\N$. Set $\varphi_n(x^1):=\varphi_0(x^1/(n+1))$
  for $x^1\in\R$ and $\phi_n:=\varphi_n\otimes\identity$ on $\Omega$.
  Then, $\phi_n\psi\in\H^1_0(\Omega)$,
  $\displaystyle\lim_{n\to\infty}\phi_n\psi=\psi$ in $\H(\Omega)$
  and  
  \begin{equation}
    H\phi_n\psi=\phi_n\:\!H\psi-2\:\!\phi_{n,1}\:\!G^{1j}\:\!\psi_{,j}
    -\phi_{n,11}\:\!G^{11}\psi-\phi_{,1}\:\!G^{1i}_{\ ,i}\:\!\psi
    \label{phi_npsi}
  \end{equation}
  in the sense of distributions.
  Using the fact that $\supp(\phi_n)$ is compact,
  Assumption~\ref{Ass1}.1 and Assumption~\ref{Ass1}.4, one
  has $\phi_n\psi\in\Dom(H)_{\rm c}$. Moreover, as a consequence
  of~(\ref{phi_npsi}) and the property
  \begin{equation*}
    \forall k\in\N\:\!,~\forall x\in\Omega\:\!,\qquad
    \partial^k_1\phi_n(x)=(n+1)^{-k}\varphi^{(k)}_0(x^1/(n+1))\:\!,
  \end{equation*}
  one also has $\displaystyle\lim_{n\to\infty}H\phi_n\psi=H\psi$ in
  $\H(\Omega)$. 
  
  \noindent
  (ii) Using point (i) and the fact that
  $\Dom(H)\subset\H^1_0(\Omega)$
  continuously and densely, one gets the following embeddings
  \begin{equation*}
    \H^1_0(\Omega)
    =\overline{\overline{\Dom(H)_{\rm c}}^{\:\!\Dom(H)}}^{\:\!\H^1_0(\Omega)}
    \subseteq\overline{\overline{\Dom(H)_{\rm c}}^{\:\!\H^1_0(\Omega)}}^{\:\!\H^1_0(\Omega)}
    =\overline{\Dom(H)_{\rm c}}^{\:\!\H^1_0(\Omega)}
    \subseteq\H^1_0(\Omega)
  \end{equation*}
  which, in particular, imply that $\Dom(H)_{\rm c}$ is dense in $\H^1_0(\Omega)$.
\end{proof}
Now, we can compute the commutator~$i\left[H,A\right]$.
\begin{Proposition}\label{Prop.commutator}
The sesquilinear form $\mathcal Q$ on $\H(\Omega)$
defined by
  \begin{equation*}
    \mathcal Q(\varphi,\psi):=i\left[\left(H\varphi,A\:\!\psi\right)
      -\left(A\:\!\varphi,H\psi\right)\right],\qquad
    \varphi,\psi\in\Dom(\mathcal Q):=\Dom(H)\cap\Dom(A)\:\!,
  \end{equation*}
  is continuous on $\Dom(H)_{\rm c}$ for the topology induced by
  $\H^1_0(\Omega)$. Moreover,
  \begin{equation}
    i\left[H,A\right]=-\partial_j\:\!G^{1j} \partial_1
    -\partial_1 G^{1j} \partial_j+\partial_i\:\!q^1 G^{ij}_{\ ,1}\partial_j
    -q^1V_{,1}\label{op. commutator}
  \end{equation}
  as operators in $\B\left(\H^1_0(\Omega),\H^{-1}(\Omega)\right)$.
\end{Proposition}
\begin{proof}
Let $\varphi,\psi\in\Dom(H)_c$.
Using the identity $A=q^1 p_1-\frac i2$ valid on
$\Dom(H)_c\subset\Dom(A)$, we have
\begin{align*}
  \mathcal Q(\varphi,\psi)
  =&\ i\left[\left(H\varphi,A\:\!\psi\right)
    -\left(A\:\!\varphi,H\psi\right)\right]
  \\
  =&\left(\varphi,H\psi\right)
  +i\left[\left(-\partial_i\:\!G^{ij}\:\!\partial_j\:\!\varphi,q^1 p_1\psi\right)
  -\left(q^1 p_1\varphi,-\partial_i\:\!G^{ij}\:\!\partial_j\:\!\psi\right)\right]\\
  &+\left(V\varphi,q^1\psi_{,1}\right)+\left(q^1\varphi_{,1},V\psi\right).
\end{align*}
In order to justify the subsequent integration by parts, we employ the
family~(\ref{regular}). Since~$\psi$ has a compact support and belongs
to~$\H_0^1(\Omega)$, it follows by using properties (iii) and~(iv) 
of Lemma~\ref{lem.regular}
that $q^1 p_1(\varepsilon)\psi\in\H_0^1(\Omega)$ for
all~$\varepsilon>0$. So, we can write
\begin{align*}
  \left(-\partial_i\:\!G^{ij}\:\!\partial_j\:\!\varphi,q^1p_1\psi\right)
  &=\lim_{\varepsilon \to 0}
  \left(-\partial_i\:\!G^{ij}\:\!\partial_j\:\!\varphi
  ,q^1p_1(\varepsilon)\:\!\psi\right)
  \\
  &=\lim_{\varepsilon \to 0}
  \left(\varphi_{,j},G^{ij}\:\!\partial_i\:\!q^1p_1(\varepsilon)\:\!\psi\right)
  \\
  &=-i\left(\varphi_{,j},G^{1j}\:\!\psi_{,1}\right)
  +\lim_{\varepsilon \to 0}
  \left(\varphi_{,i},G^{ij}q^1 p_1(\varepsilon)\:\!\psi_{,j}\right)
\end{align*}
and similarly for the integral
\begin{equation*}
  \left(q^1p_1\:\!\varphi,-\partial_i\:\!G^{ij}\:\!\partial_j\:\!\psi\right)
  =i\left(\varphi_{,1},G^{1j}\:\!\psi_{,j}\right)
  +\lim_{\varepsilon \to 0}
  \left(p_1(\varepsilon)^*\varphi_{,i},q^1G^{ij}\:\!\psi_{,j}\right).
\end{equation*}
Since
\begin{align*}
  \lim_{\varepsilon\to0}
  \left(p_1(\varepsilon)^*\varphi_{,i},q^1G^{ij}\:\!\psi_{,j}\right)
  =&\ \lim_{\varepsilon\to0}
  \left(\varphi_{,i},p_1(\varepsilon)\:\!q^1\:\!G^{ij}\:\!\psi_{,j}\right)
  \\
  =&\ -i\big[\left(\varphi_{,i},G^{ij}\:\!\psi_{,j}\right)
  +\big(\varphi_{,i},q^1G^{ij}_{\ ,1}\:\!\psi_{,j}\big)\big]
  \\
  &\ +\lim_{\varepsilon\to0}
  \left(\varphi_{,i},q^1G^{ij} p_1(\varepsilon)\:\!\psi_{,j}\right),
\end{align*}
and
\begin{equation*}
  \left(q^1\varphi_{,1},V\psi\right)
  =-\left(\varphi,\partial_1\:\!q^1 V\psi\right)
  =-\left(\varphi,V\psi\right)
  -\left(\varphi,q^1 V_{,1}\:\!\psi\right)
  -\left(\varphi,q^1 V\psi_{,1}\right),
\end{equation*}
we finally obtain that
\begin{equation}
  \mathcal Q(\varphi,\psi)
  =\left(\varphi_{,j},G^{1j}\:\!\psi_{,1}\right)
  +\left(\varphi_{,1},G^{1j}\:\!\psi_{,j}\right)
  -\big(\varphi_{,i},q^1G^{ij}_{\ ,1}\psi_{,j}\big)
  -\left(\varphi,q^1 V_{,1}\psi\right).\label{form commutator}
\end{equation}
This implies that $\mathcal Q$ restricted to $\Dom(H)_{\rm c}$
is continuous for the topology induced by $\H^1_0(\Omega)$.
Now, $\Dom(H)_{\rm c}$ is dense in $\H^1_0(\Omega)$
by Lemma~\ref{density}.(ii). Thus, $\mathcal Q$ defines (by
continuous extension) an operator in $\B(\H_0^1(\Omega),\H^{-1}(\Omega))$,
which we shall denote $i\left[H,A\right]$. Furthermore,
using~(\ref{form commutator}), we obtain~(\ref{op. commutator})
in $\B(\H_0^1(\Omega),\H^{-1}(\Omega))$.
\end{proof}
%
\subsubsection{Strict Mourre estimate for the free Hamiltonian}
%
Now we prove that~$H_0$ is of class~$C^\infty(A)$ and~$A$ strictly
conjugate to it on~$\R\setminus\mathcal T$. So, let us first recall
the following definition \cite[Sec.~7.2.1\,\&\,7.2.2]{ABG}\:\!:
\begin{Definition}
  Let~$A,\:\!H$ be self-adjoint operators
  in a Hilbert space~$\H$ with~$H$ of class~$C^1(A)$.
  Furthermore, if~$S,\:\!T\in\B(\H)$, we write~$S\gtrsim T$
  if there exists~$K\in\K(\H)$ so that~$S\geq T+K$.
  Then,~$\forall\lambda\in\R$\:\!,
  \begin{align*}
    &\varrho^A_H(\lambda):=\sup\left\{a\in\R\,:\,\exists\:\!
    \varepsilon>0~~\textrm{s.t.}~~E^H(\lambda;\varepsilon)\:\!i\left[H,A\right]
    E^H(\lambda;\varepsilon)\geq a\:\!E^H(\lambda;\varepsilon)\right\},\\
    &{\widetilde\varrho}^A_H(\lambda):=\sup\left\{a\in\R\,:\,\exists\:\!
    \varepsilon>0~~\textrm{s.t.}~~E^H(\lambda;\varepsilon)\:\!i\left[H,A\right]
    E^H(\lambda;\varepsilon)\gtrsim a\:\!E^H(\lambda;\varepsilon)\right\}
  \end{align*}
where~$E^H(\lambda;\varepsilon):= E^H((\lambda-\varepsilon,\lambda+\varepsilon))$
designates the spectral projection of~$H$ for the interval
$(\lambda-\varepsilon,\lambda+\varepsilon)$.
\end{Definition}
We also need the following natural generalisation of~\cite[Thm.~3.4]{BG}.
\begin{Theorem}\label{Ck}
Let~$H_1,\:\!H_2$ be two self-adjoint, bounded from below operators in the
Hilbert spaces~$\H_1,\:\!\H_2$. 
Assume that $A_\jmath$\:\!,~$\jmath=1,2$, is a self-adjoint operator
in~$\H_\jmath$ such that~$H_\jmath$ is of class~$C^k(A_\jmath)$,
$k\in\left(\N\setminus\{0\}\right)\cup\{+\infty\}$.
Let $H:= H_1\otimes\identity+\identity\otimes H_2$ and 
$A:= A_1\otimes\identity+\identity\otimes A_2$, which are  
self-adjoint operators in $\H_1\otimes\H_2$. Then~$H$ is of class~$C^k(A)$ and
$\forall\lambda\in\R$\:\!\emph{:}
\begin{equation*}
  \varrho^A_H(\lambda)=\inf_{\lambda=\lambda_1+\lambda_2}
  \left[\varrho^{A_1}_{H_1}(\lambda_1)+\varrho^{A_2}_{H_2}(\lambda_2)\right].
\end{equation*}
\end{Theorem}

\begin{Corollary}\label{Prop.reg.H0} $H_0\in C^\infty(A)$ and
\begin{equation}
  \forall\lambda\in\R\:\!,\qquad\varrho^A_{H_0}(\lambda)
  =\begin{cases}
  2\:\!\rho(\lambda) & \textrm{if}\quad\lambda\geq\nu_1\\
  +\infty & \textrm{if}\quad\lambda<\nu_1\:\!,
  \end{cases}\label{mourre H_0}
\end{equation}
where~$\rho(\lambda):=\lambda-\sup\left\{\zeta\in\mathcal T\,:\,\zeta\leq\lambda\right\}$
is strictly positive on~$\R\setminus\mathcal T$.

\end{Corollary}
\begin{proof}
  $A_1:=A_\shortparallel$, $A_2:=0$ are self-adjoint in~$\sii(\R)$, 
  respectively~$\sii(\omega)$.
  $H_1:= p^2_1$, $H_2:=-\Delta^\omega_{\rm D}$ are self-adjoint, bounded from below
  in~$\sii(\R)$, respectively~$\sii(\omega)$. Clearly,
  \cite[Ex.~6.2.8]{ABG}~$p^2_1\in C^\infty(A_\shortparallel)$ 
  and~$-\Delta^\omega_{\rm D}\in C^\infty(0)$.
  The first part of the claim and~(\ref{mourre H_0}) then follows from Theorem~\ref{Ck}.
  The expression for~$\rho(\lambda)$ is a direct consequence of the respective behaviours of
  \cite[Sec.~7.2.1]{ABG}~$\varrho^{A_\shortparallel}_{p^2_1}$ and~$\varrho^0_{-\Delta^\omega_{\rm D}}$\:\!:
  \begin{equation*}
    \begin{bmatrix}
      \varrho^{A_\shortparallel}_{p^2_1}(\lambda_1)\vspace{5pt}\\ \varrho^0_{-\Delta^\omega_{\rm D}}(\lambda_2)
    \end{bmatrix}
    =\begin{cases}
    \begin{bmatrix} 2\lambda_1 \\ +\infty\end{bmatrix}
      & \textrm{if} \quad\begin{bmatrix}\lambda_1\geq0\\\lambda_1<0\end{bmatrix}\\\\
      \begin{bmatrix} 0 \\+\infty\end{bmatrix}
	& \textrm{if} \quad\begin{bmatrix} \lambda_2\in\mathcal T
	\\ \lambda_2\in\R\setminus\mathcal T\end{bmatrix}.
    \end{cases}
  \end{equation*}
\end{proof}
%
\subsubsection{Regularity of the Hamiltonian}
%
In order to prove the regularity of~$H$, we need two technical lemmas.
\begin{Lemma}\label{Lemma.4.3}
  $\forall z\in\R\setminus\sigma(H)$\:\!, $\forall\vartheta\leq 1$\:\!,
  one has
  \begin{itemize}
  \item[\emph{(i)}] $\left[R(z),\langle q^1\rangle^\vartheta\right]
    \in\B\left(\H(\Omega),\H_0^1(\Omega)\right)$\:\!,
  \item[\emph{(ii)}] $\forall i\in\{1,\ldots,d\}$\:\!,~~$\left[R(z),
    \langle q^1\rangle^\vartheta\right]
    \partial_i\in\B\left(\H(\Omega),\H_0^1(\Omega)\right)$\:\!.
  \end{itemize}
\end{Lemma}
\noindent
This is established by adapting the proof of~\cite[Lemma 4.3]{Benbernou2}
while next Lemma follows from the use of~\cite[Proof~of~Prop.~4.2]{Benbernou2}.
\begin{Lemma}\label{Lemma.theta} 
  Let~$S\in\B\left(\H(\Omega)\right)$
  be self-adjoint and~$\vartheta\in(0,1]$\:\!, then
    \begin{equation*}
      \langle q^1\rangle^\vartheta S\in\B\left(\H(\Omega),
      \H^\vartheta(\R)\otimes \sii(\omega)\right)
      \ \Longrightarrow\ S\in C^\vartheta(A)\:\!.
    \end{equation*}
\end{Lemma} 
\noindent
(Note that the proof involves principally two facts.
First, $S\in\B\left(\H(\Omega),\Dom(|A|^\vartheta)\right)$ implies
$S\in C^\vartheta(A)$. Second, the continuous imbedding
$\H_\vartheta^\vartheta(\R)\subseteq\Dom(|A_\shortparallel|^\vartheta)$,
which follows by real interpolation~\cite[Sec.~2.7]{ABG} from
the continuous imbedding $\H_1^1(\R)\subseteq\Dom(|A_\shortparallel|)$.)
\begin{Remark}\label{H C1}
  The facts that~$i\left[H,A\right]\in\B\left(\H^1_0(\Omega),\H^{-1}(\Omega)\right)$
  and that~$\H^1_0(\Omega)$
  is stable under $\{\e^{iAt}\}_{t\in\R}$ imply
  \emph{\cite[Sec.~6.3]{ABG}} that $H\in C^1(A)$.
\end{Remark}
\begin{Proposition}\label{Prop.reg.H}
$\exists\:\!\vartheta\in(0,1]$~~such that~~$H\in C^{1+\vartheta}(A)$\:\!.
\end{Proposition}
\begin{proof}
We show that each term appearing in the expression
for $B:=i\left[H,A\right]$ is at least of class
$C^\gamma(A)$ for a certain $\gamma\in(0,1]$.

Consider first
$B_1:=-\partial_j\:\!G^{j1}\:\!\partial_1-\partial_1\:\!G^{1j}\:\!\partial_j$.
An explicit calculation (analogous to that of the proof of
Proposition~\ref{Prop.commutator}) implies that
\begin{equation*}
  i\left[B_1,A\right]
  =-2\:\!\partial_1 G^{11}\:\!\partial_1
  -\partial_1G^{1j}\:\!\partial_j
  - \partial_j\:\!G^{j1}\:\!\partial_1
  + \partial_j\:\!q^1G^{j1}_{\ ,1}\:\!\partial_1
  + \partial_1 q^1 G^{1j}_{\ ,1}\:\!\partial_j
\end{equation*}
as operators in~$\B(\H_0^1(\Omega),\H^{-1}(\Omega))$.
Thus, $B_1\in C^1(A)$ by Remark~\ref{H C1}. 

Let $z\in\R\setminus\sigma(H)$. As a consequence of the fact that
$H\in C^1(A)$, one can interpret $i\left[A,R(z)\right]$ as the product
of \cite[Sec.~6.2.2]{ABG} three bounded operators, 
\viz\/ $R(z):\H(\Omega)\to\Dom(H)$, $B:\Dom(H)\to\Dom(H)^*$ and
$R(z):\Dom(H)^*\to\H(\Omega)$. Thus, using Proposition
\ref{Prop.commutator}, one can write as an operator
identity in $\B\left(\H(\Omega)\right)$
\begin{align*}
  i\left[A,R(z)\right]
  = R(z)\:\!B\:\!R(z)
  =&\ R(z)\:\!B_1\:\!R(z)
  +R(z)\:\!\partial_i\:\!q^1G^{ij}_{\ ,1}\:\!\partial_jR(z)
  \\
  &-R(z)\:\!q^1V_{,1}\:\!R(z)\:\!.
\end{align*}
Since the first term has already been shown to be bounded,
it is enough to prove that the second and third terms
on the r.h.s. are of class~$C^\gamma(A)$ for some~$\gamma\in(0,1]$.

We employ Lemma~\ref{Lemma.theta} with
$\vartheta:=\min\{\vartheta_1,\vartheta_2\}$ in order to deal with
both terms. Using some commutation relations, we get
\begin{align*}
  \langle q^1\rangle^\vartheta
  R(z)\:\!\partial_i\:\!q^1 G^{ij}_{\ ,1} \partial_j\:\!R(z)
  =&\ R(z)\:\!\partial_i\:\!\langle q^1\rangle^\vartheta
  q^1 G^{ij}_{\ ,1}\partial_j\:\!R(z)
  \\
  &-\left[R(z),\langle q^1\rangle^\vartheta\right]
  \partial_i\:\!q^1 G^{ij}_{\ ,1} \partial_j\:\!R(z)
  \\
  &-R(z) \left[\partial_i,\langle q^1\rangle^\vartheta\right]
  q^1 G^{ij}_{\ ,1} \partial_j\:\!R(z)\:\!.
\end{align*}
Under Assumption~\ref{Ass1}.3, the first term on the r.h.s. is
in~$\B(\H(\Omega),\H_0^1(\Omega))$.
The second and the last one are
in~$\B(\H(\Omega),\H_0^1(\Omega))$
by Lemma~\ref{Lemma.4.3}.(ii) and the boundedness
of~$\langle q^1 \rangle^\vartheta_{,1}$, respectively. Moreover,
\begin{equation*}
  \langle q^1 \rangle^\vartheta R(z)\:\!q^1 V_{,1}\:\!R(z)
  = R(z)\langle q^1 \rangle^\vartheta q^1 V_{,1}\:\!R(z)
  +\left[\langle q^1 \rangle^\vartheta,R(z)\right] q^1 V_{,1}\:\!R(z)
\end{equation*}
is in~$\B\left(\H(\Omega),\H_0^1(\Omega)\right)$
by Assumption~\ref{Ass2}.3 and Lemma~\ref{Lemma.4.3}.(i). Thus, all
the terms in the expression of $B$ are at least of class
$C^\vartheta(A)$. This implies the claim.
\end{proof}
%

\subsubsection{The main result}\label{Sec.Main}
%
\begin{Proposition}\label{Prop.conj.H}
  $\forall\lambda\in\R\setminus\threshold$,~~$\widetilde
  \varrho_{H}^A(\lambda)>0$\:\!.
\end{Proposition}
\begin{proof}
  Corollary~\ref{Prop.reg.H0} and Proposition~\ref{Prop.reg.H} imply
  that both~$H_0$ and~$H$ are of class $C_u^1(A)$.
  Furthermore,~$R(i)-R_0(i)$ is compact by Proposition~\ref{Prop.compact},
  with the result that $\widetilde\varrho_H^A=\widetilde\varrho_{H_0}^A$
  due to~\cite[Thm.~7.2.9]{ABG}.
  Finally, since~\cite[Prop.~7.2.6]{ABG}~$\widetilde\varrho_{H_0}^A\geq\varrho^A_{H_0}$,
  we can conclude using~Corollary~\ref{Prop.reg.H0}.
\end{proof}
Summing up, we result in the following spectral properties of~$H$.
\begin{Theorem}\label{Thm.main}
Let~$\omega$ be a bounded open connected
set in~$\R^{d-1}$,~$d \geq 2$,
and denote by~$\threshold$
the set of eigenvalues of~$-\Delta_{\rm D}^\omega$.
Let~$H$ be the operator~\emph{(\ref{hamiltonian})}
with~$\Omega:=\R \times \omega$, subject to Dirichlet boundary
conditions, and satisfying \emph{Assumptions~\ref{Ass1}} and~\emph{\ref{Ass2}}.
Then
\begin{itemize}
\item[\emph{(i)}]
$\sigma_\mathrm{ess}(H)=\left[\kappa,\infty\right)$,
where~$\kappa:=\inf\threshold$,
\item[\emph{(ii)}]
$\sigma_\mathrm{sc}(H)=\varnothing$\:\!,
\item[\emph{(iii)}]
$\sigma_\mathrm{p}(H) \cup \threshold$
is closed and countable\:\!,
\item[\emph{(iv)}]
$\sigma_\mathrm{p}(H) \setminus \threshold$ is composed of
finitely degenerated eigenvalues, which can accumulate
at the points of~$\threshold$ only\:\!,
\end{itemize}
\end{Theorem}
\begin{proof}
The claim~(i) is included in Proposition~\ref{Prop.compact}.
Since~$A$ is conjugate to~$H$ on \mbox{$\R\setminus\threshold$}
by Proposition~\ref{Prop.conj.H},
the assertions (ii)--(iv) follow by the abstract
conjugate operator method~\cite[Thm.~7.4.2]{ABG}.  
\end{proof}
To conclude this section, let us remark that
Assumptions~\ref{Ass1}.3 and~\ref{Ass2}.3 could be weakened.
Firstly, we recall that the situation with~$V=0$
and~$G= \rho\:\!\identity$, $\rho$~being a real-valued function greater than
a strictly positive constant, 
is investigated in~\cite{Benbernou2,DDI} where the authors
admit local singularities of~$\rho$.
More specifically, one assumes that
$\rho=\rho_{\rm s}+\rho_\ell$,
where~$\rho_\ell$ is the part satisfying a condition
analogous to Assumption~\ref{Ass1}.3, 
while~$\rho_{\rm s}$ need not be differentiable.
(In~\cite{Benbernou2},~$\supp(\rho_{\rm s})$ is assumed to be compact.
The result of~\cite{DDI} is better in the sense that~$\rho_s$
is only supposed to be a short-range perturbation there.
However, this requires strengthening of the condition
analogous to Assumption~\ref{Ass1}.2 about the decay of~$\rho$ at infinity.)
Secondly, the optimal conditions one has to impose
on the potential of a Schr\"odinger operator
are known~\cite{BGM,ABG}.
\setcounter{equation}{0}
\section{Curved tubes}\label{Sec.tubes}
%
In this part, we use Theorem~\ref{Thm.main} in order
to find geometric sufficient conditions which guarantee
that the spectral results of the theorem hold true 
for curved tubes.  
%
\subsection{Geometric preliminaries}
%
\subsubsection{The reference curve}
%
Given~$d \geq 2$, let~$p:\R \to \R^d$ be a regular unit-speed
smooth (\ie,~$C^\infty$-smooth) curve satisfying the following hypothesis.
\begin{Assumption}\label{Ass.Frenet}
There exists a collection of~$d$ smooth mappings
$
  e_i : \R \to \R^d
$
with the following properties\:\!:
\begin{enumerate}
\item 
$\forall i,j \in\{1,\dots,d\}$,~$\forall
s\in\R$\:\!,~~$e_{i}(s)\cdot e_{j}(s) = \delta_{ij}$\:\!,
\item
$\forall i \in\{1,\dots,d-1\}$,~$\forall s\in\R$\:\!,~~the~$i^{\rm th}$
derivative~$p^{(i)}(s)$ of~$p(s)$ lies in the span
of~$e_1(s),\dots,e_i(s)$\:\!,
\item 
$
  e_1 = \dot p
$\:\!,
\item 
$\forall s\in\R$\:\!,~~$\{e_1(s),\dots,e_d(s)\}$ has the positive orientation\:\!,
\item 
$\forall i \in\{1,\dots,d-1\}$,~$\forall s\in\R$\:\!,~~$\dot e_i(s)$
lies in the span of~$e_1(s),\dots,e_{i+1}(s)$\:\!.
\end{enumerate}
\end{Assumption}
\noindent
Here and in the sequel, 
``\,$\cdot$\,''~denotes the inner product in~$\R^d$.
\begin{Remark}
A vector field with the property~1 is called
a \emph{moving frame} along~$p$
and it is a \emph{Frenet frame} if 
it satisfies~2 in addition, \cf~\emph{\cite[Sec.~1.2]{Kli}}\:\!.
A sufficient condition to ensure the existence
of the frame of \emph{Assumption~\ref{Ass.Frenet}}
is to require that~\emph{\cite[Prop.~1.2.2]{Kli}}\:\!,
for all~$s \in \R$\:\!, the vectors
$
  \dot{p}(s), p^{(2)}(s), \dots, p^{(d-1)}(s)
$
are linearly independent.  
This is always satisfied if $d=2$.
However, we do not assume \emph{a priori}
the above non-degeneracy condition for $d \geq 3$
because it excludes the curves such that, for some open
$I\subseteq\R$\:\!, $p\!\upharpoonright\! I$ lies in a
lower-dimensional subspace of~$\R^d$.
\end{Remark}
\noindent
The properties of~$\{e_1, \dots, e_d\}$
summarised in Assumption~\ref{Ass.Frenet}
yield~\cite[Sec.~1.3]{Kli} the Serret-Frenet formulae,
\begin{equation}\label{Frenet}
  \dot{e}_i =\K_{i}^{\ j} \:\! e_j
\end{equation}
with~$\K \equiv\big(\K_i^{\ j}\big)$ being a skew-symmetric~$d \times d$ matrix
defined by
\begin{equation}\label{curvature}
  \K :=
  \begin{pmatrix}
   0                & \kappa_1 &               & \textrm{\LARGE 0}\\
   -\kappa_1        & \ddots   & \ddots        &                 \\
                    & \ddots   & \ddots        & \kappa_{d-1}    \\
   \textrm{\LARGE 0} &          & -\kappa_{d-1} & 0
   \end{pmatrix}.
\end{equation}
Here~$\kappa_i$ is called the~$i^\mathrm{th}$~\emph{curvature} of~$p$.
Under our Assumption~\ref{Ass.Frenet}, the curvatures are smooth functions
of the arc-length parameter~$s\in\R$.

\subsubsection{The appropriate moving frame}\label{Sec.rot}
%
In this subsection, we introduce another moving frame along~$p$,
which better reflects the geometry of the curve,
and will be used later to define a tube about it. 
We shall refer to it as the \emph{Tang frame} 
because it is a natural generalisation
of the Tang frame known from the theory of $3$-di\-men\-sional
waveguides~\cite{Tsao-Gambling_1989,GJ,DE}.
Our construction follows the generalisation introduced in~\cite{CDFK}.

Let the~$(d-1)\times(d-1)$ matrix~$(\mathcal R_{\mu}^{\ \nu})$ be defined by
the system of differential equations
\begin{equation}\label{diff.eq}
\dot{\mathcal R}_{\mu}^{\ \nu} + \mathcal R_{\mu}^{\ \alpha} \:\! \K_{\alpha}^{\ \nu} = 0
\end{equation}
with~$(\mathcal R_{\mu}^{\ \nu}(s_0))$ being a rotation matrix
in~$\R^{d-1}$ for some $s_0 \in \R$ as initial condition, \ie,
\begin{equation}\label{rotation}
  \det\left(\mathcal R_{\mu}^{\ \nu}(s_0)\right)=1
  \qquad\textrm{and}\qquad
  \delta_{\alpha\beta}\:\!\mathcal R_\mu^{\ \alpha}(s_0)\:\!
  \mathcal R_\nu^{\ \beta}(s_0)=\delta_{\mu\nu}\:\!.
\end{equation}
The solution of~(\ref{diff.eq}) exists and is smooth 
by standard arguments in the theory of differential equations
(\cf~\cite[Sec.~4]{Kurzweil}). Furthermore, the conditions~(\ref{rotation})
are satisfied for \emph{all} $s_0\in\R$. Indeed, by means of Liouville's
formula~\cite[Thm.~4.7.1]{Kurzweil} and~$\tr(\K)=0$, one checks
that~$\det\left(\mathcal R_{\mu}^{\ \nu}\right)=1$ identically, while
the validity of the second condition for all $s_0\in\R$ is obtained via
the skew-symmetry of~$\K$\:\!:
\begin{equation*}
\left(\delta_{\alpha\beta}\mathcal R_\mu^{\ \alpha}
\mathcal R_\nu^{\ \beta}\right)^{\boldsymbol.}
=-\mathcal R_\mu^{\ \alpha}\big(\delta_{\gamma\beta}\K_\alpha^{\ \gamma}
+\delta_{\alpha\gamma}\K_\beta^{\ \gamma}\big)\:\!\mathcal R_\nu^{\ \beta}=0\:\!.
\end{equation*}

We set
\begin{equation*}
  \mathcal R \equiv \big(\mathcal R_{i}^{\ j}\big)
  :=
  \begin{pmatrix}
    1 & 0 \\
    0 & \big(\mathcal R_{\mu}^{\ \nu}\big)
  \end{pmatrix}
\end{equation*}
and introduce the Tang frame as the moving frame 
$
  \{\tilde e_1, \dots, \tilde e_d\}
$
along~$p$ defined by
\begin{equation}\label{Tang rotations}
  \tilde e_i := \mathcal R_{i}^{\ j} \:\! e_j\:\!.
\end{equation}
Combining~(\ref{Frenet}) with~(\ref{diff.eq}), one easily finds
\begin{equation}\label{Frenet.bis}
  \dot{\tilde e}_1 = \kappa_1 \:\! e_2
  \qquad\textrm{and}\qquad
  \dot{\tilde e}_\mu = \mathcal R_\mu^{\ \alpha}\K_\alpha^{\ 1}e_1
  = - \kappa_1 \:\! \mathcal R_{\mu}^{\ 2} \:\! e_1\:\!.
\end{equation}

The interest of the Tang frame will appear in the
following subsection.
%
\subsubsection{The tube}\label{SubSec.tubes}
%
Let~$\omega$ be a bounded open connected set in~$\R^{d-1}$.
Without loss of generality, we assume that~$\omega$
is translated so that its centre of mass is at the origin.
Let~$\Omega:=\R\times\omega$ be a straight tube.
We define the curved tube~$\strip$ of the same cross-section~$\omega$
about~$p$ as the image of the mapping
\begin{equation}\label{tube.map}
  \mathscr L: \Omega \to \R^d\:\!,\qquad
  \left(s,u^2,\dots,u^d\right) \mapsto p(s) + \tilde e_\mu(s) \:\! u^\mu\:\!,
\end{equation}
\ie,~$\strip:=\mathscr L(\Omega)$.  

As already mentioned in Introduction, the shape of the curved tube~$\strip$  
of cross-section~$\omega$ about~$p$ depends on the choice of
rotations~$(\mathcal R_{\mu}^{\ \nu})$ in~(\ref{Tang rotations}),
unless~$\omega$ is rotation invariant. As usual in the theory of quantum
waveguides (see, \eg,~\cite{DE,CDFK}), we restrict ourselves to the technically
most advantageous choice determined by~(\ref{diff.eq}), \ie,~when the
cross-section~$\omega$ rotates along~$p$ w.r.t.\ the Tang frame (another choice
can be found in~\cite{EFK}).

We write $u\equiv(u^2,\dots,u^d)$, define $a:=\sup_{u\in\omega}|u|$ and always assume 
\begin{Assumption}\label{Ass.basic}
\rule{0.5\textwidth}{0ex}
\begin{enumerate}
\item $\kappa_1\in\sinf(\R)$ and~$a\:\!\|\kappa_1\|_\infty< 1$\:\!,
\item $\strip$ does not overlap itself\:\!.
\end{enumerate}
\end{Assumption}
\noindent
Then, the mapping~$\mathscr L:\Omega\to\strip$ is a diffeomorphism. Indeed,
by virtue of the inverse function theorem,
the first condition guarantees that it is a local diffeomorphism
which is global through the injectivity induced by the second condition.
Consequently,~$\mathscr L^{-1}$ determines a system of global
(\emph{geodesic} or \emph{Fermi}) ``coordinates''~$(s,u)$.
At the same time, the tube~$\strip$ can be identified
with the Riemannian manifold~$(\Omega,g)$,
where~$g\equiv(g_{ij})$ is
the metric tensor induced by the immersion~(\ref{tube.map}),
that is~$g_{ij}:=\mathscr L_{,i}\cdot\mathscr L_{,j}$. 
The formulae~(\ref{Frenet.bis}) yield
\begin{equation}\label{metric}
  g =\diag\left(h^2,1,\ldots,1\right)
  \qquad\textrm{with}\qquad
  h(s,u) := 1+u^\mu\:\!\mathcal R_\mu^{\ \alpha}(s)\:\!\K_\alpha^{\ 1}(s)\:\!.
\end{equation}
Note that the metric tensor~(\ref{metric}) is diagonal due to our special
choice of the ``transverse'' frame $\{\tilde{e}_2,\dots,\tilde{e}_d\}$,
which is the advantage of the Tang frame.

We set~$|g| := \det(g) = h^2$, which defines through
$\de v:=h(s,u)\:\!\de s\:\!\de u$ the volume element of~$\strip$;
here~$\de u$ denotes the~$(d-1)$-dimensional Lebesgue measure
in~$\omega$.   
\begin{Remark}[Low-dimensional examples]
When~$d=2$, the cross-section~$\omega$ is just the interval~$(-a,a)$,
the curve~$p$ has only one curvature~$\kappa:=\kappa_1$,
the rotation matrix~$(\mathcal R_{\mu}^{\ \nu})$ equals
(the scalar)~$1$ and
\begin{equation*}
  h(s,u)=1-\kappa(s)\:\!u\:\!.
\end{equation*}
If~$d=3$, it is convenient to make the Ansatz
\begin{equation*}
  \left(\mathcal R_\mu^{\ \nu}\right) =
  \begin{pmatrix}
    \cos\alpha & -\sin\alpha \\
    \sin\alpha & \cos\alpha
  \end{pmatrix},
\end{equation*}
where~$\alpha$ is a real-valued differentiable function.
Then, it is easy to see that~\emph{(\ref{diff.eq})} 
reduces to the differential equation~$\dot\alpha=\tau$,
where~$\tau$ is the torsion of~$p$,
\ie,~one puts~$\kappa:=\kappa_1$ and~$\tau:=\kappa_2$.
Choosing~$\alpha$ as an integral of~$\tau$, we can write
\begin{equation*}
  h(s,u) =
  1-\kappa(s)\left[u^2\cos\alpha(s)+u^3\sin\alpha(s)\right].
\end{equation*}
\end{Remark}
\begin{Remark}[On Assumption~\ref{Ass.basic}]
If~$p$ were a compact embedded curve, then
\emph{Assumption~\ref{Ass.basic}} could always be achieved for
sufficiently small~$a$. In general, however, one cannot
exclude self-intersections of the tube using the local geometry of
an embedded curve~$p$ only. One way to avoid this disadvantage would
be to consider~$(\Omega,g)$ as an abstract Riemannian manifold where
only the curve~$p$ is embedded in~$\R^d$. Nonetheless, in the present
paper, we prefer to assume \emph{Assumption~\ref{Ass.basic}.2}
\emph{a priori}  because $\strip$~does not have a physical meaning if
it is self-intersecting. Finding global geometric conditions on~$p$
ensuring the validity of \emph{Assumption~\ref{Ass.basic}.2} is an
interesting question, which is beyond the scope of the present paper, however. 
\end{Remark}
%

\subsection{The Laplacian}\label{Sec.laplacian}
%
Our object of interest is the Dirichlet Laplacian~(\ref{laplacian}),
with~$\strip$ defined by~(\ref{tube.map}). We construct it as follows.
Using the diffeomorphism~(\ref{tube.map}),
we identify the Hilbert space $\sii(\strip)$
with~$\sii(\Omega,\de v)$
and consider on the latter the Dirichlet form
\begin{equation}\label{form}
  \widetilde Q(\varphi,\psi) :=
  \int_{\Omega} \overline{\varphi_{,i}}\:\!g^{ij}\psi_{,j}
  \:\!\de v\:\!,
  \qquad
  \varphi,\psi\in\Dom(\widetilde Q) :=
  \H_0^1(\Omega,\de v)\:\!,
\end{equation}
where~$(g^{ij}):= g^{-1}$.
The form~$\widetilde Q$ is clearly densely defined, non-negative, symmetric
and closed on its domain. Consequently, there exists a unique
non-negative self-adjoint operator~$\widetilde H$
satisfying~$\Dom(\widetilde H)\subset\Dom(\widetilde Q)$
associated with~$\widetilde Q$.We have
\begin{align}\label{hamiltonian.tilde}
  \widetilde H\psi&=-|g|^{-1/2}\partial_i\:\!|g|^{1/2} g^{ij}\partial_j\:\!\psi\:\!,\\
  \psi\in\Dom(\widetilde H)&=
  \big\{\psi\in\H_0^1(\Omega,\de v) \,:\,
  \partial_i\:\!|g|^{1/2} g^{ij} \partial_j\:\!\psi
  \in\sii(\Omega,\de v) \big\}.
\end{align}
That is,~$\widetilde H$ is the Laplacian~(\ref{laplacian})
expressed in the coordinates~$(s,u)$.

In order to apply Theorem~\ref{Thm.main}, 
we transform~$\widetilde H$ into a unitarily equivalent
operator~$H$ of the form~(\ref{hamiltonian})
acting on the Hilbert space~$\H(\Omega):=\sii(\Omega)$,
without the additional weight~$|g|^{1/2}$ in the volume element.
This is achieved by means of the unitary mapping
$\mathcal U:\sii(\Omega,\de v)\to\H(\Omega),\psi \mapsto |g|^{1/4}\psi$.
Defining~$H := \mathcal U\widetilde H\:\!\mathcal U^{-1}$, one has
\begin{align}\label{hamiltonian.tube}
  H\psi&=
  -|g|^{-1/4}\partial_i\:\!|g|^{1/2} g^{ij}
  \partial_j |g|^{-1/4}\:\!\psi\:\!,\\
  \psi\in\Dom(H)&=
  \big\{\psi\in\H_0^1(\Omega) \,:\,
  \partial_i\:\!|g|^{1/2}g^{ij}\partial_j\:\!|g|^{-1/4}\:\!\psi
  \in\sii(\Omega) \big\}.
\end{align}
Commuting~$|g|^{-1/4}$ with the gradient components
in the expression for~$H$, we obtain on~$\Dom(H)$
\begin{equation}\label{hamiltonian.potential}
  H = -\partial_i\:\!g^{ij} \partial_j + V\:\!,
\end{equation}
where
\begin{equation}\label{potential}
  V :=
  - \frac54 \:\!\frac{(h_{,1})^2}{h^4}
  + \frac12 \:\! \frac{h_{,11}}{h^3}
  - \frac14 \:\! \frac{\delta^{\mu\nu}\:\!h_{,\mu}\:\!h_{,\nu}}{h^2}
  + \frac12 \:\! \frac{\delta^{\mu\nu}\:\!h_{,\mu\nu}}{h}\:\!.
\end{equation}
Actually, (\ref{hamiltonian.potential}) with~(\ref{potential})
is a general formula valid for any smooth metric
of the form~$g=\diag(h^2,1,\dots,1)$.
In our special case with~$h$ given by~(\ref{metric}),
we find that
$h_{,\mu\nu}=0$,
$
  \delta^{\mu\nu} h_{,\mu}h_{,\nu}
  =\delta^{\alpha\beta}
  \K_\alpha^{\ 1}\K_\beta^{\ 1}
$ 
by~(\ref{rotation}),
while~(\ref{diff.eq}) gives
\begin{align}\label{diff.metric1}
  h_{,1}(\cdot,u)&=u^\mu\:\!\mathcal R_\mu^{\ \alpha}
  \big(\dot\K_\alpha^{\ 1}
  -\K_\alpha^{\ \beta}\K_\beta^{\ 1}\big)\:\!,
  \\
  h_{,11}(\cdot,u)&=u^\mu\:\!\mathcal R_\mu^{\ \alpha}
  \big(\ddot\K_\alpha^{\ 1}
  -\dot\K_\alpha^{\ \beta}\K_\beta^{\ 1}
  -2\:\!\K_\alpha^{\ \beta}\dot\K_\beta^{\ 1}
  +\K_\alpha^{\ \beta}\K_\beta^{\ \gamma}\K_\gamma^{\ 1}
  \big)\:\!.\nonumber
\end{align}
%
\subsection{Results}\label{Sec.result.tubes}
%
It remains to impose decay conditions on the curvatures of~$p$
(and their derivatives) in order that the
operator~(\ref{hamiltonian.potential})
satisfies Assumption~\ref{Ass1} and Assumption~\ref{Ass2}.

Let us first consider the more general situation
where
the matrix~$(g^{ij})$ is equal to~$\diag(h^{-2},1,\dots,1)$
with the explicit dependence of~$h$ on~$s$ and~$u$
not specified. One shows that it is sufficient to impose
the following hypotheses.
\begin{Assumption}\label{Ass}
Uniformly for $u\in\omega$\:\!,
\begin{enumerate}
\item $h(s,u)\longrightarrow1$~~as~~$|s|\to\infty$\:\!,
\item 
$h_{,11}(s,u)$,~$(\delta^{\mu\nu}
\:\!h_{,\mu}h_{,\nu})(s,u)$,~$\delta^{\mu\nu}\:\!h_{,\mu\nu}(s,u)
\longrightarrow0$~~as~~$|s|\to\infty$\:\!,
\item $\exists\:\!\vartheta\in(0,1]$~~s.t.
\begin{equation*}
  h_{,1}(s,u),~h_{,111}(s,u),~(\delta^{\mu\nu}
  \:\!h_{,\mu}h_{,\nu})_{,1}(s,u),~\delta^{\mu\nu}\:\!h_{,1\mu\nu}(s,u)
  =\mathcal{O}\big(|s|^{-(1+\vartheta)}\big)\:\!.
\end{equation*}
\end{enumerate}
\end{Assumption}
\noindent
Indeed, the first hypothesis supplies Assumption~\ref{Ass1}.2,
while Assumption~\ref{Ass1}.1 is fulfilled due to
basic Assumption~\ref{Ass.basic}.
Next, since~$h$ is a smooth function,
Assumption~\ref{Ass}.2 together with the behaviour of~$h_{,1}$
in Assumption~\ref{Ass}.3 are sufficient to ensure
both Assumption~\ref{Ass2}.1 and Assumption~\ref{Ass2}.2.
It is also clear that the asymptotic behaviour of~$h_{,1}$
in Assumption~\ref{Ass}.3 supplies Assumption~\ref{Ass1}.3.
Assumption~\ref{Ass1}.4 holds true due to Assumption~\ref{Ass1}.3  
and the particular form of~$(g^{ij})$.
It remains to check Assumption~\ref{Ass2}.3.
This is easily done by calculating
the derivative of the potential~(\ref{potential})\:\!:
\begin{align*}
  V_{,1}
  = 5&\:\!\frac{(h_{,1})^3}{h^5}
  - 4\:\!\frac{h_{,1} \:\! h_{,11}}{h^4}
  + \frac{h_{,111}}{2\:\!h^3}\\
  &+ \frac{\delta^{\mu\nu}}2\left(
  \frac{h_{,1}\:\!h_{,\mu}\:\!h_{,\nu}}{h^3}
  - \frac{h_{,1}\:\!h_{,\mu\nu} + \:\!h_{,1\mu}\:\!h_{,\nu}}{h^2}
  + \frac{h_{,1\mu\nu}}h
  \right).
\end{align*}
With~$h$ given by~(\ref{metric}), we find in addition
to~(\ref{diff.metric1}) that~$h_{,1\mu\nu}=0$ and
\begin{align*}
  (\delta^{\mu\nu}h_{,\mu}h_{,\nu})_{,1}
  =&\ 2\:\!\delta^{\alpha\beta}\dot\K_\alpha^{\ 1}\K_\beta^{\ 1}
  \\
  h_{,111}(\cdot,u)=&\ u^\mu\mathcal R_\mu^{\ \alpha}\big(
  \dddot\K_\alpha^{\ 1}
  -\ddot\K_\alpha^{\ \beta}\K_\beta^{\ 1}
  -3\:\!\K_\alpha^{\ \beta}\ddot\K_\beta^{\ 1}
  -3\:\!\dot\K_\alpha^{\ \beta}\dot\K_\beta^{\ 1}
  +\dot\K_\alpha^{\ \beta}\K_\beta^{\ \gamma}\K_\gamma^{\ 1}
  \\
  &\phantom{\ u^\mu\mathcal R_\mu^{\ \alpha}\big(}
  +2\:\!\K_\alpha^{\ \beta}\dot\K_\beta^{\ \gamma}\K_\gamma^{\ 1}
  +3\:\!\K_\alpha^{\ \beta}\K_\beta^{\ \gamma}\dot\K_\gamma^{\ 1}
  -\K_\alpha^{\ \beta}\K_\beta^{\ \gamma}\K_\gamma^{\ \delta}
  \K_\delta^{\ 1}
  \big)\:\!.
\end{align*}
Since~$|u^\mu\mathcal R_\mu^{\ \alpha}| < a$,
Assumption~\ref{Ass} holds true provided we impose
the following conditions on the curvatures
\begin{Assumption}\label{Ass.curvature} 
\rule{0.5\textwidth}{0ex}
\begin{enumerate}
\item 
$\forall\alpha\in\{2,\dots,d\}$,~~$\K_\alpha^{\ 1}(s),~\ddot
\K_\alpha^{\ 1}(s)\longrightarrow0$~~as~~$|s|\to\infty$\:\!,
\item 
$\forall\alpha,\beta\in\{2,\dots,d\}$,~~$\K_\alpha^{\ \beta},~\dot
\K_\alpha^{\ 2}\in L^\infty(\R)$\:\!,
\item 
$\exists\:\!\vartheta\in(0,1]$~~s.t.~~$\forall\alpha\in\{2,\dots,d\}$,
\begin{equation*}
  \dot\K_\alpha^{\ 1}(s)
  ,~\dddot\K_\alpha^{\ 1}(s),~\K_\alpha^{\ 2}(s)
  ,~\ddot\K_\alpha^{\ 2}(s)
  ,~\big(\dot\K_\alpha^{\ \beta}\K_\beta^{\ 2}\big)(s)
  ,~\big(\K_\alpha^{\ \beta}\dot\K_\beta^{\ 2}\big)(s)
  =\mathcal{O}\big(|s|^{-(1+\vartheta)}\big)\:\!.
\end{equation*}
\end{enumerate}
\end{Assumption}
\begin{Remark}
  These conditions reduce to those of \emph{Theorem~\ref{Thm}} provided~$d=2$.
  When $d=3$, it is sufficient to assume 
  the conditions of \emph{Theorem~\ref{Thm}} for the first curvature,
  and $\dot\kappa_2\in L^\infty(\R)$
  and $\kappa_2(s),\ddot{\kappa}_2(s)
  =\mathcal{O}\left(|s|^{-(1+\vartheta)}\right)$
  for some $\vartheta\in(0,1]$\:\!.
\end{Remark}

We conclude this section by applying Theorem~\ref{Thm.main}.
\begin{Theorem}\label{Thm.tubes}
Let~$\strip$ be a tube defined via~\emph{(\ref{tube.map})}
about a smooth infinite curve embedded in~$\R^d$.
Suppose \emph{Assumptions~\ref{Ass.Frenet}},
\emph{\ref{Ass.basic}} and~\emph{\ref{Ass.curvature}}.
Then all the spectral results \emph{(i)--(iv)} of \emph{Theorem~\ref{Thm.main}}
hold true for the Dirichlet Laplacian on~$\sii(\strip)$.
\end{Theorem}

\setcounter{equation}{0}
\section{Curved strips on surfaces}\label{Sec.strips}

In this final section, we investigate the situation
where the ambient space is a general Riemannian manifold
instead of the Euclidean space~$\R^d$.
We restrict ourselves to $d=2$,
\ie,~$\strip$ is a strip around an infinite curve
in an (abstract) two-dimensional surface.
We refer to~\cite{K1} for basic spectral properties
of~$-\Delta^\Gamma_{\rm D}$ and geometric details.

\subsection{Preliminaries}\label{Sec.Def}

Let~$\mathcal{A}$ be a smooth connected complete non-compact
two-dimensional Riemannian manifold of
bounded Gauss curvature~$K$.
Let \mbox{$p: \R\to\mathcal{A}$}
be a smooth unit-speed curve embedded in~$\mathcal{A}$
with (geodesic) curvature~$\kappa$ and denote by
\mbox{$n: \R\to T_{p(\cdot)}\mathcal{A}$}
a smooth unit normal vector field along~$p$.
Given~$a>0$, we consider the straight strip
$
  \Omega := \R\times(-a,a)
$
and define a curved strip~$\strip$ of same width over~$p$
as a~$a$-tubular neighbourhood of~$p$ in~$\mathcal{A}$ by
\begin{equation}\label{exp}
  \strip :=\mathscr L(\Omega)\:\!,
  \qquad\textrm{where}\qquad
  \mathscr L: (s,u) \mapsto
  \exp_{p(s)}(u\:\!n(s))\:\!.
\end{equation}
Note that~$s\mapsto\mathscr L(s,u)$ traces the curves
parallel to~$p$ at a fixed distance~$|u|$,
while the curve~$u\mapsto\mathscr L(s,u)$ is a unit-speed
geodesic orthogonal to~$p$ for any fixed~$s$.
We always assume
\begin{Assumption}\label{Ass.diff}
 ~$\mathscr L: \Omega \to \strip$~~is a diffeomorphism,
\end{Assumption}
\noindent
Then~$\mathscr L^{-1}$
determines a system of Fermi ``coordinates''~$(s,u)$,
\ie,~the geodesic coordinates based on~$p$.
The metric tensor of~$\Gamma$ in these coordinates
acquires~\cite[Sec.~2.4]{Gray} the diagonal form
\begin{equation}\label{s.metric}
  g(s,u) = \diag\left(h^2(s,u),1\right),
\end{equation}
where~$h$ is a smooth function satisfying the Jacobi equation
\begin{equation}\label{Jacobi}
  h_{,22}+K\:\!h=0
  \qquad\textrm{with}\qquad\left\{
  \begin{aligned}
    &h(\cdot,0)=1\\
    &h_{,2}(\cdot,0)=-\kappa\:\!.
  \end{aligned}\right.
\end{equation}
Here~$K$ and~$\kappa$ are considered as functions of the Fermi coordinates
(the sign of~$\kappa$ being uniquely determined
up to the re-parameterisation~$s \mapsto -s$
or the choice of~$n$).
The determinant of the metric tensor,~$|g|:=\det(g)=h^2$,
defines through~$\de v := h(s,u)\:\!\de s\:\!\de u$ the area
element of the strip.

Assuming that the metric~$g$
is uniformly elliptic in the sense that
\begin{Assumption}\label{Ass.elliptic}~$\exists\:\! c_\pm
  \in(0,\infty)$~~s.t.~~$\forall(s,u)\in\Omega$,~~$c_- \leq h(s,u) \leq c_+$
\end{Assumption}
\noindent
holds true, the Dirichlet Laplacian corresponding to~$\strip$ can
be defined in the same way as in Section~\ref{Sec.laplacian}, \ie,~as
the operator~$\widetilde H$ associated with the form~(\ref{form}),
satisfying~(\ref{hamiltonian.tilde}). At the same time, we may introduce
the unitarily equivalent operator~$H$ on~$\sii(\Omega)$ given
by~(\ref{hamiltonian.tube}) and satisfying~(\ref{hamiltonian.potential})
with~(\ref{potential}).
\begin{Remark}
If \emph{Assumption~\ref{Ass.elliptic}} holds true, then the inverse
function theorem together with~\emph{(\ref{Jacobi})} yield that
\emph{Assumption~\ref{Ass.diff}} is satisfied for all sufficiently
small~$a$ provided the strip~$\strip$ does not overlap itself.
\emph{Assumption~\ref{Ass.elliptic}} is satisfied, for instance,
if~$\strip$ is a sufficiently thin strip on a ruled surface,
\cf~\emph{\cite[Sec.~7]{K1}}.
\end{Remark}
%

\subsection{Results}

In view of the more general approach
in the beginning of Section~\ref{Sec.result.tubes},
we see that Assumption~\ref{Ass} (with~$d=2$) guarantees
Assumptions~\ref{Ass1} and~\ref{Ass2} also in the present case.
Applying Theorem~\ref{Thm.main}, we obtain,
with~$\threshold=\{n^2 \nu_1\}_{n=1}^\infty$
where~$\nu_1:=\pi^2/(2a)^2$, the following result

\begin{Theorem}\label{Thm.general}
Let~$\strip$ be a tubular neighbourhood of radius~$a>0$
about a smooth infinite curve,
which is embedded in a smooth connected complete
non-compact surface of bounded curvature. 
Suppose
\emph{Assumptions~\ref{Ass.diff}}, \emph{\ref{Ass.elliptic}} and
\emph{\ref{Ass}}. Then all the spectral results \emph{(i)--(iv)} of
\emph{Theorem~\ref{Thm.main}} hold true for the Dirichlet Laplacian
on~$\sii(\strip)$.
\end{Theorem}
Assume now that the strip is \emph{flat} in the sense of~\cite{K1},
\ie,~the curvature~$K$ is equal to zero everywhere on~$\strip$.
Then the Jacobi equation~(\ref{Jacobi}) has the explicit solution
(\cf~(\ref{metric}) for~$d=2$)
\begin{equation}\label{FlatStrip}
  h(s,u)=1-\kappa(s)\:\!u
\end{equation}
and Assumption~\ref{Ass} can be replaced by some conditions
on the decay of the curvature~$\kappa$ at infinity,
namely, we adopt Assumption~\ref{Ass.curvature}
with~$\kappa_1 \equiv \kappa$ and $\mathcal{K}_\mu^{\ \nu} = 0$
(\cf~the assumptions of Theorem~\ref{Thm}).
At the same time, it easy to see that Assumption~\ref{Ass.diff}
and~\ref{Ass.elliptic} are satisfied if
Assumption~\ref{Ass.basic} holds true.
\begin{Theorem}[Flat strips]\label{Thm.flat}
Let~$\strip$ be a tubular neighbourhood of radius \mbox{$a>0$}
about a smooth infinite curve of curvature~$\kappa$,
which is embedded in a smooth connected complete non-compact
surface of bounded curvature~$K$ 
such that~$K\:\!\upharpoonright\:\!\strip = 0$.
Suppose \emph{Assumption~\ref{Ass.basic}} and
\begin{enumerate}
\item $\kappa(s)$,~$\ddot{\kappa}(s)$ 
$\longrightarrow0$~~as~~$|s|\to\infty$\:\!,
\item $\exists\:\!\vartheta\in(0,1]$~~s.t.~~$\dot{\kappa}(s)$,~$\dddot{\kappa}(s)$
$=\mathcal{O}\left(|s|^{-(1+\vartheta)}\right)$\:\!.
\end{enumerate}
Then, all the spectral results \emph{(i)--(iv)} of \emph{Theorem~\ref{Thm.main}}
hold true for the Dirichlet Laplacian on~$\sii(\strip)$.
\end{Theorem}
%

\section*{Acknowledgements}

\addcontentsline{toc}{section}{Acknowledgements} One of the authors (D.K.)
would like to thank M.~M\v{a}ntoiu and R.~Purice for proposing to him to use
the conjugate operator method in the theory of quantum waveguides and for
many discussions on the topic. We are very grateful to W. Amrein  
for his careful reading of the present paper
and S. Richard for his useful suggestions and comments.
This work was partially supported by FCT/POCTI/FEDER, Portugal,
GA AS CR grant IAA 1048101 and the Fonds National Suisse.
%
%
\addcontentsline{toc}{section}{References}
\providecommand{\bysame}{\leavevmode\hbox to3em{\hrulefill}\thinspace}

\end{document}